# A transfer learning metamodel using artificial neural networks applied to natural convection flows in enclosures

Majid Ashouri[1] and Alireza Hashemi[2]

[1]College of Engineering, Boston University, Boston, MA 02215, USA. ashouri@bu.edu

[2]The Decision Lab, Montreal, QC H4C 2C7, Canada. ahashemi@thedecisionlab.com

**Abstract**

In this paper, we employed a transfer learning technique to predict the Nusselt number for natural convection flows in enclosures. Specifically, we considered the benchmark problem of a two-dimensional square enclosure with isolated horizontal walls and vertical walls at constant temperatures. The Rayleigh and Prandtl numbers are sufficient parameters to simulate this problem numerically. We adopted two approaches to this problem: Firstly, we made use of a multi-grid dataset in order to train our artificial neural network in a cost-effective manner. By monitoring the training losses for this dataset, we detected any significant anomalies that stemmed from an insufficient grid size, which we further corrected by altering the grid size or adding more data. Secondly, we sought to endow our metamodel with the ability to account for additional input features by performing transfer learning using deep neural networks. We trained a neural network with a single input feature (Rayleigh) and extended it to incorporate the effects of a second feature (Prandtl). We also considered the case of hollow enclosures, demonstrating that our learning framework can be applied to systems with higher physical complexity, while bringing the computational and training costs down.

## 1. Introduction

An exhaustive metamodel using machine learning and particularly artificial neural networks (ANNs) can be used to characterize the flow and heat transfer behaviors of an engineering system [1-10]. Data-driven ANN metamodels are especially useful for predicting heating or cooling processes in natural convection flows. Natural convection occurs when temperature gradients, and subsequently, density differences in a fluid induce buoyancy effects. This phenomenon has applications in various engineering systems such as nuclear reactors, heat exchangers, solar energy collectors, and electronic devices. The strength of natural convection can be quantified by the Nusselt number (Nu). Several ANN metamodels for predicting Nu in natural convection systems have been investigated using numerical simulations [11-14] or by conducting experiments [14-18].

As the complexity of a physical system increases, the training of an ANN metamodel becomes more challenging. In this scenario, creating a suitable training dataset may become laborious and expensive, thus, one often has to make a compromise between cost and accuracy. Transfer learning (TL) is a common approach that can be employed to mitigate this issue by using knowledge from a source domain to improve the learning of a target domain with limited data. One approach to TL is to use a multi-fidelity training procedure to improve model accuracy by applying higher-fidelity training data to a model trained on low-fidelity data. This method was shown to effectively reduce training costs in previous studies [19-27].

One can conceive of several case examples for the application of TL in heat transfer systems: The need for TL may arise when the performance of a heat exchanger is affected due to fouling over time, or TL can be applied to construct a semi-empirical metamodel using readily available simulation data and limited experimental data. TL can also enable us to use information from a low-cost two-dimensional metamodel to account for three-dimensional effects using a small three-dimensional dataset. Moreover, TL is especially beneficial for domain adaptation [28], where information from a different (but related) heat transfer system can be applied to solve problems in another system (e.g., heat transfer from vertical and horizontal walls can be related to heat transfer from an enclosure).

The key parameter characterizing a natural convection problem is the Rayleigh number (Ra). This parameter is defined as the product of the Grashof number (the ratio of buoyancy to viscous forces), and Prandtl number (Pr, the ratio of momentum and thermal diffusivity). However, various expected and unexpected factors can affect a natural convection system (i.e., some features may become active under different conditions, or after a system is redesigned). For example, the system's orientation with respect to gravity or the system's boundary conditions may change, or applying a magnetic force to the system could lead to thermomagnetic [29,30] or magnetohydrodynamic [31,32] effects. Moreover, generating a new set of simulations or experiments can be a laborious process. Domain adaptation or TL with deep neural networks (DNN) can be employed to remedy these problems.

The objective of the present study is to develop a TL framework for building Nu metamodels for natural convection flows in enclosures. First, we demonstrate how to exploit a seemingly imperfect dataset generated by cost-effective simulations by using different grid systems. Second, we build a Nu metamodel by constructing a DNN for transfer learning. We consider the benchmark problem of an air-filled enclosure and train our neural network to predict Nu for different Ra inputs. We transfer the learning of this model to also consider enclosures filled with an arbitrary fluid. Finally, to demonstrate the applicability of the present TL framework for more complex natural convection problems, we considered the case of hollow enclosures and looked at whether our TL approach could incorporate a third input parameter.

## 2. Problem Description

We aimed to extract a metamodel out of a physical model that numerically predicts the natural convection characteristics of a square enclosure filled with a Newtonian fluid. This problem is governed by two parameters: Ra and Pr (see Appendix A for details). We consider Ra of up to $10^8$ and Pr of greater than 0.05 ($1 < \text{Ra} \leq 10^8$ and $0.05 \leq \text{Pr} < \infty$); however, lower Pr were also considered provided that the ratio of Ra/Pr is less than $10^8$.

A 400×400 grid system was shown to provide precise results for Nu even for the most stringent cases. Appendix B provides details related to the numerical method and grid independence test. Using a single logical processor on a 2.6 GHz Intel Core i7-3720QM CPU, an average computational time of about 4,850 seconds (as high as about 13,000 seconds for low Pr) was spent on obtaining the numerical solutions using a 400×400 grid system. Nonetheless, as demonstrated in Appendix C, we demonstrate that lower grid systems can provide accurate numerical solutions for limited ranges of Ra. For example, a 200×200 grid system (with an average simulation time of 1,300 seconds) can reliably be used for Ra of up to $10^7$ with errors of less than 0.5%. Therefore, we consider a multi-grid simulation that also uses lower grid systems, wherever possible, to decrease the simulation cost in training our metamodel.

## 3. Results and Discussion

The multi-grid dataset that we used for training is shown in a scatter graph (Fig. 1a). This dataset includes a limited number of simulation data points using a 400×400 grid system (5% of the data) for cases with high Ra or low Pr. The rest of this dataset includes numerical results using 200×200 (9%), 50×50 (15%), and 25×25 grids (71%), for Ra within the range of $10^5 \leq \text{Ra} \leq 10^7$, $10^3 < \text{Ra} < 10^5$, and $1 < \text{Ra} \leq 10^3$, respectively. Nonetheless, we also included simulations of higher fidelity solutions that were carried out as part of the analysis in Appendix C. As can be seen in Fig. 1a, using low-cost simulations (25×25 grid systems, with an average computational time of about 40 seconds) allowed us to generate more data in the low Ra region, in order to capture the nonlinear variation of Nu for Ra $\lesssim 10^3$. In contrast, for higher Ra, Nu varied in a logarithmically linear manner (Fig. C-1a).

An optimized ANN (as described in Appendix D) was used to construct a metamodel for predicting Nu. We trained our ANN using 480 data points (in which 15% of the dataset was considered for validation during training) according to the details presented in Table 1 under "step 1". The contour graph for the relative errors associated with the training and validation datasets used for this model is presented in Fig. 1b.

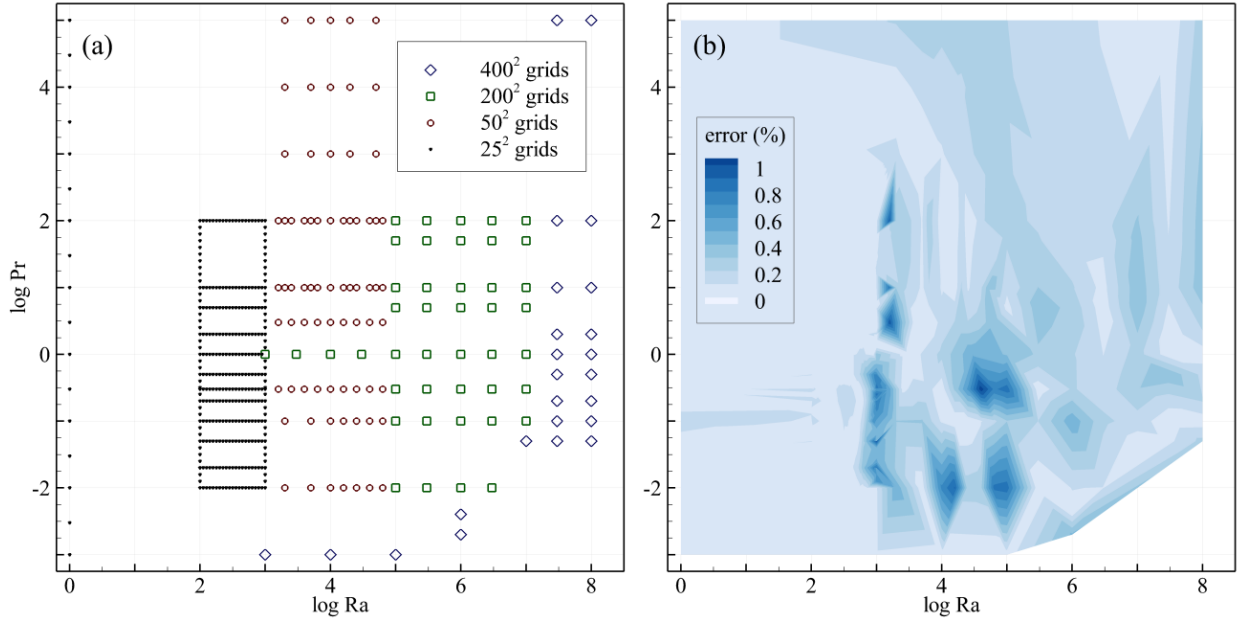

Fig. 1. a) The scatter graph for the training dataset generated by numerical simulations using a combination of 25×25, 50×50, 200×200, and 400×400 grid systems. b) The error contour for the training dataset.

The error contour in Fig. 1b can be used to determine if the grid system for each data point was properly selected. We assumed that the source of any significant deviance in the error values in Fig. 1b was due to a lack of sufficient data points or possibly an inconsistency between the results of different grid systems. Considering these two possibilities, we revised our training dataset by adding more data points within regions with high Ra or low Pr, and replaced some of our data with higher fidelity simulation results (Fig. 2a). We fed the new dataset to our previously trained ANN to achieve an improved validation loss (by 48%) as summarized in Table 1 under "step 2". The contour plot of the relative errors for the training and validation datasets is shown in Fig. 2b.

Table 1. The details of the ANN training for the prediction of Nu.

| | Training data | | | | Training error | | Validation error | | Model cost | |
|---|---|---|---|---|---|---|---|---|---|---|
| Step No. | $25^2$ | $50^2$ | $200^2$ | $400^2$ | MSE ($/10^{-6}$) | MAE ($/10^{-3}$) | MSE ($/10^{-6}$) | MAE ($/10^{-3}$) | ST (hr) | NNT (hr) |
| 1 | 340 | 73 | 43 | 24 | 1.12 | 0.64 | 1.77 | 0.86 | 53.7 | 2.0 |
| 2 | 339 | 68 | 65 | 36 | 1.03 | 0.60 | 0.93 | 0.57 | 79.6[*] | 0.8 |

MSE: mean squared error loss; MAE: mean absolute error; ST: simulation time; NNT: ANN training time.
[*] Also includes the simulation time for the data that was removed from the original dataset.

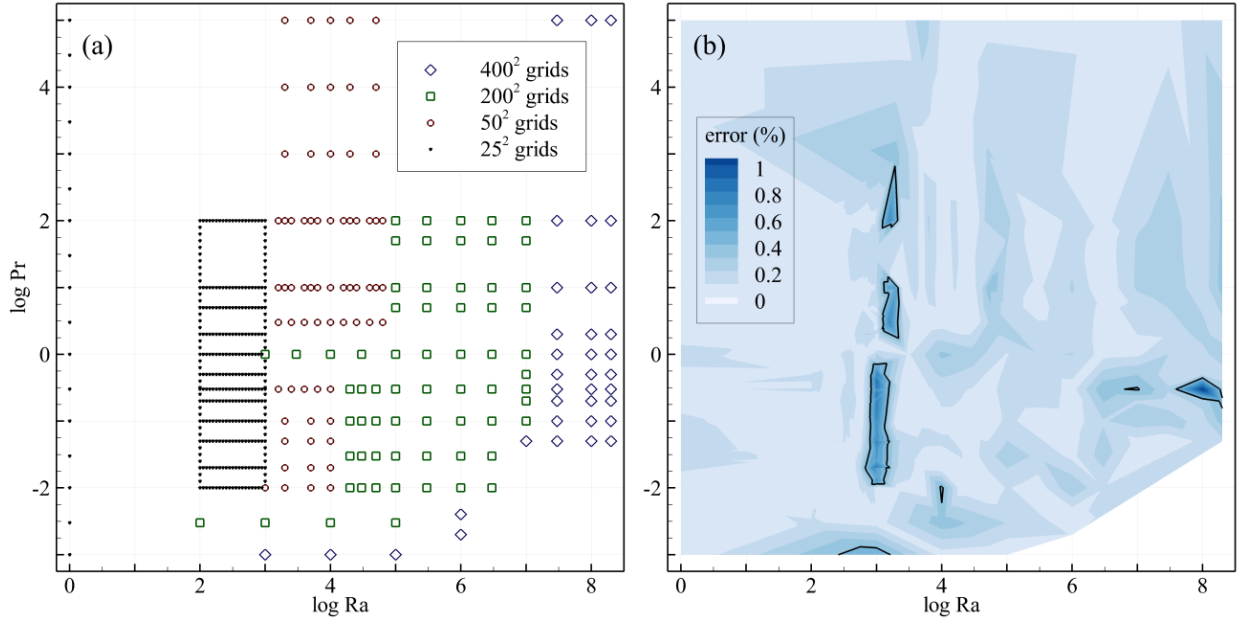

Fig. 2. a) The scatter graph for the revised training dataset. b) The relative error contour for the training dataset. The line contour surrounds areas with errors greater than 0.4%.

We tested our ANN metamodel using a test dataset of 100 simulations using a 400×400 grid system. Our test points were randomly drawn from normal distributions of $\log(Ra) = 4.8 \pm 2.1$ and $\log(Pr) = 0 \pm 1.7$. A scatter graph for the test dataset is presented in Fig. 3a. The test result for our ANN model is summarized in Table 2. There was an improvement of 49% in the test loss (MSE) between the two steps at an added simulation cost of 46%. The test error contour for the final metamodel is presented in Fig. 3b. This metamodel predicts Nu with an error of $0.22 \pm 0.21$ % (the first and second terms are the mean and standard deviation of the relative errors, respectively). In Fig. 3b, the highest errors are for low Pr. As before, one can revise the training dataset based on the error contour of Fig. 2b to achieve higher accuracy.

Table 2. The details of the test result for the Nu metamodel.

| Step No. | MSE ($/10^{-6}$) | MAE ($/10^{-3}$) | MRE (%) | SD (%) |
|---|---|---|---|---|
| 1 | 3.48 | 1.19 | 0.27 | 0.33 |
| 2 | 1.77 | 0.96 | 0.22 | 0.21 |

MSE: mean squared error loss; MAE: mean absolute error;
MRE: mean relative error; SD: standard deviation of relative error.

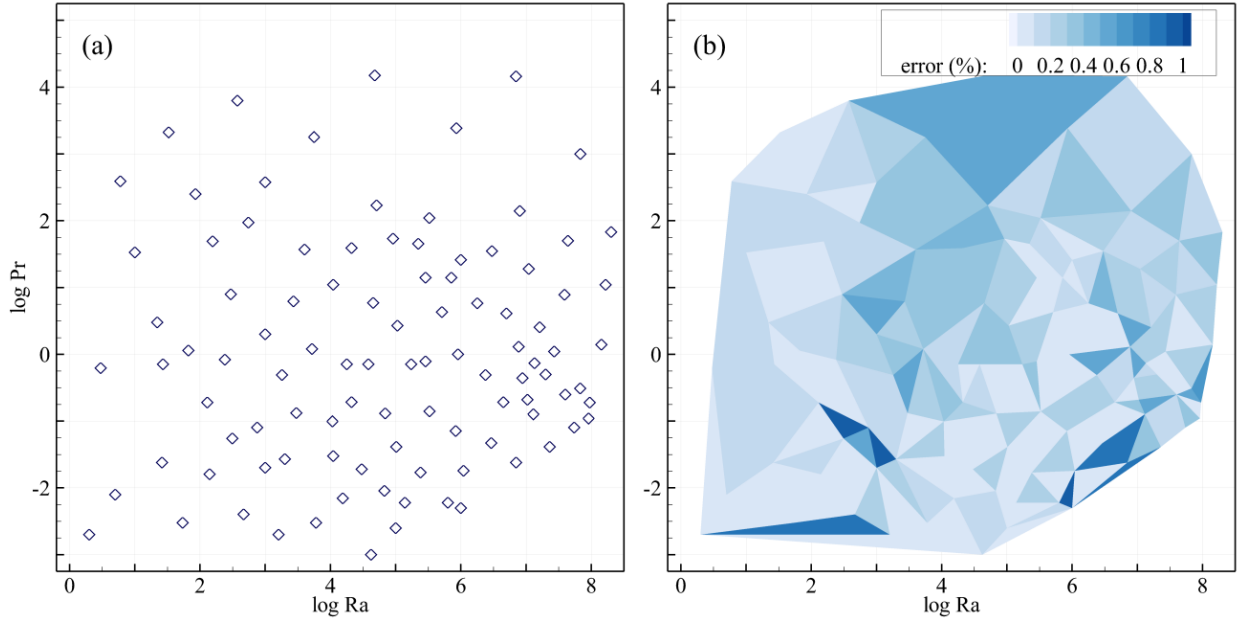

Fig. 3. a) The scatter graph for the test dataset generated by numerical simulations using 400×400 grids. b) The relative error contour for the test dataset. The highest relative error is 0.97%.

As an alternative to the above approach, we developed a TL framework that enabled us to incorporate any potential feature. For this approach, we constructed a metamodel based on one input (in this case, Ra) and later added additional input parameters (namely, Pr).

For simplicity, we only made use of 200×200 and 400×400 grid systems for our simulations. We initially generated a metamodel that predicted the variation of Nu with Ra for an air-filled enclosure (Pr = 0.71). We trained an ANN as illustrated in Fig. 4 (as block I) using 30 data points (ranging from Ra = 1 to $2 \times 10^8$), without using a validation dataset during training (with the goal of lowering simulation costs). The results of testing our Ra-Nu metamodel using 10 test data points is presented in Table 3 under Step 1.

We extended our metamodel to also consider the effects of Pr. We applied the same structure as Ra (the right branch in Fig. 4) and merged the outputs of the Pr branch and block I using a Multiply layer (Keras API). We added a one-node layer after the Multiply layer to adjust for the multiplication coefficient. We froze the layers on block I, and trained the new hidden layers using data points from Step 1 as well as a new dataset of constant Ra simulation points (24 training data at a fixed Ra = $10^5$ and variable Pr ranging from Pr = $10^{-3}$ to $10^5$). We trained the ANN using 54 data points and validated it using 20 data points. The test results (using the dataset of Fig. 3a) are presented in Table 3 under Step 2.

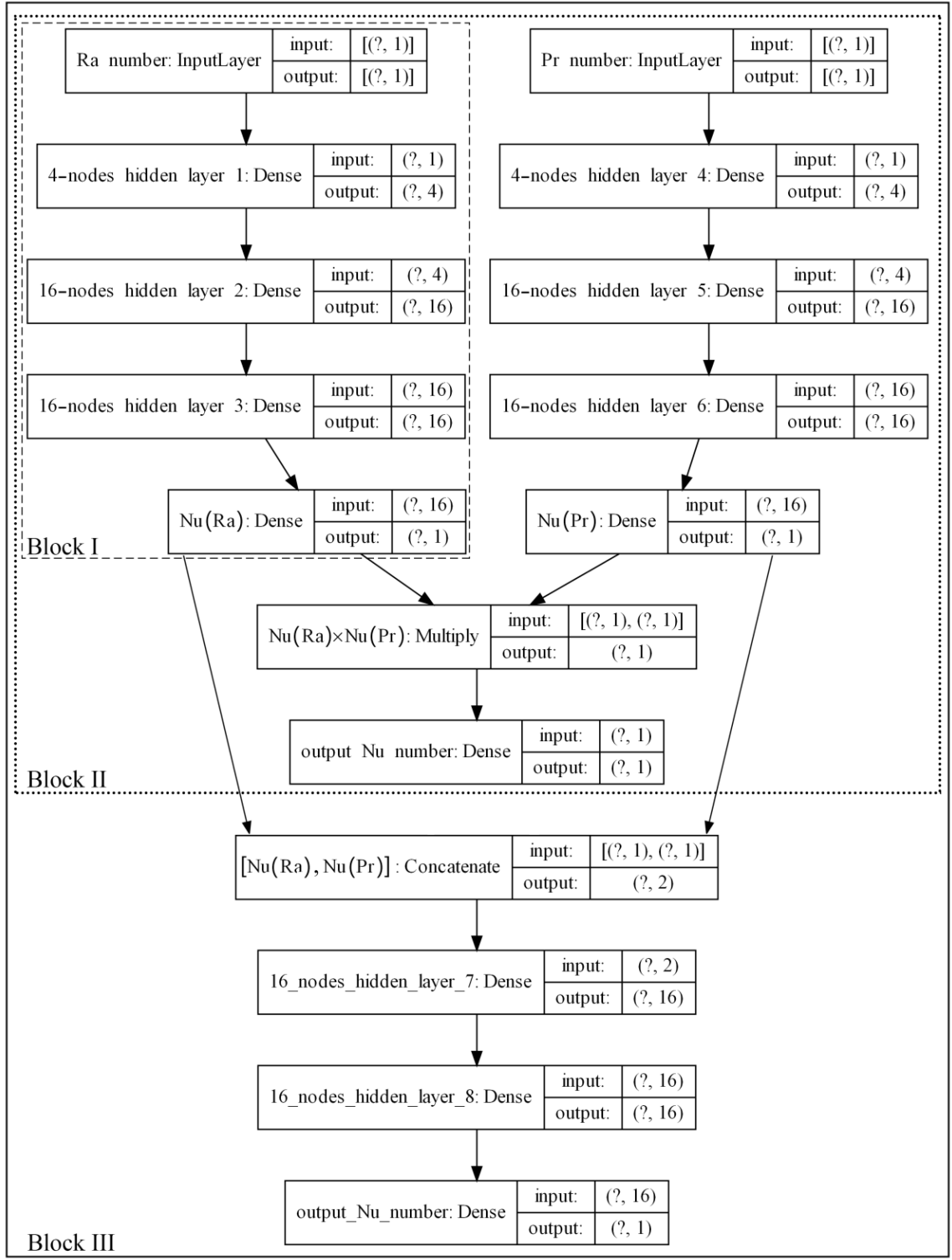

Fig. 4. The structure of the DNN for transfer learning. Block I consists of a metamodel that predicts the variation of Nu with Ra (where Pr is unchanged). In Block II, there is a second branch that takes Pr into account as a new input. The outputs of the two branches are Multiplied to produce the output of Block II. In Block III, the outputs are concatenated and two hidden layers are added to the model in order to improve the training.

Table 3. The details of the DNN training for the prediction of Nu in a square enclosure.

| Step No. | Training data | | ANN parameters | | Test error | | | Model cost | |
|---|---|---|---|---|---|---|---|---|---|
| | $200^2$ | $400^2$ | Trainable | Non-trainable | MSE (/$10^{-5}$) | MAE (/$10^{-3}$) | RE** (%) | ST (hr) | NNT (hr) |
| 1: Block I | 23 | 7 | 377 | – | 0.01 | 0.30 | 0.07±0.04 | 19.2* | 0.2 |
| 2: Block II | 42 | 32 | 379 | 377 | 22.2 | 9.35 | 2.12±2.55 | 42.7 | 0.8 |
| 3: Block III | 64 | 53 | 337 | 754 | 2.39 | 3.08 | 0.71±0.89 | 75.7 | 3.6 |

* Also includes the simulation time for the test data that was generated for this step.
** Mean ± standard deviation of relative error.

To enhance accuracy, one can remove the Multiply layer from block II and replace it with a concatenated layer followed by two additional hidden layers (Block III in Fig. 4). We froze the training on the two branches in Fig. 4 and applied the weight values from step 2 (Block II). We trained the added hidden layers in Block III using new simulation points at fixed $Ra = 10^8$ or at fixed $Pr = 0.05$ on top of the previous dataset. As presented in Table 3 (Step 3), our results on the test dataset improved remarkably after applying this procedure. The scatter graph for the different training datasets that we employed in this approach is presented in Fig. 5a. The test error contour for this approach is shown in Fig. 5b (for Step 3). Further improvements in model accuracy can be achieved after retraining the DNN from step 3 using more data. For example, using 31 additional data points (shown by grey symbols in Fig. 5a) decreased the test loss by 51% (RE = 0.45±0.66) at an increased simulation cost of 28%.

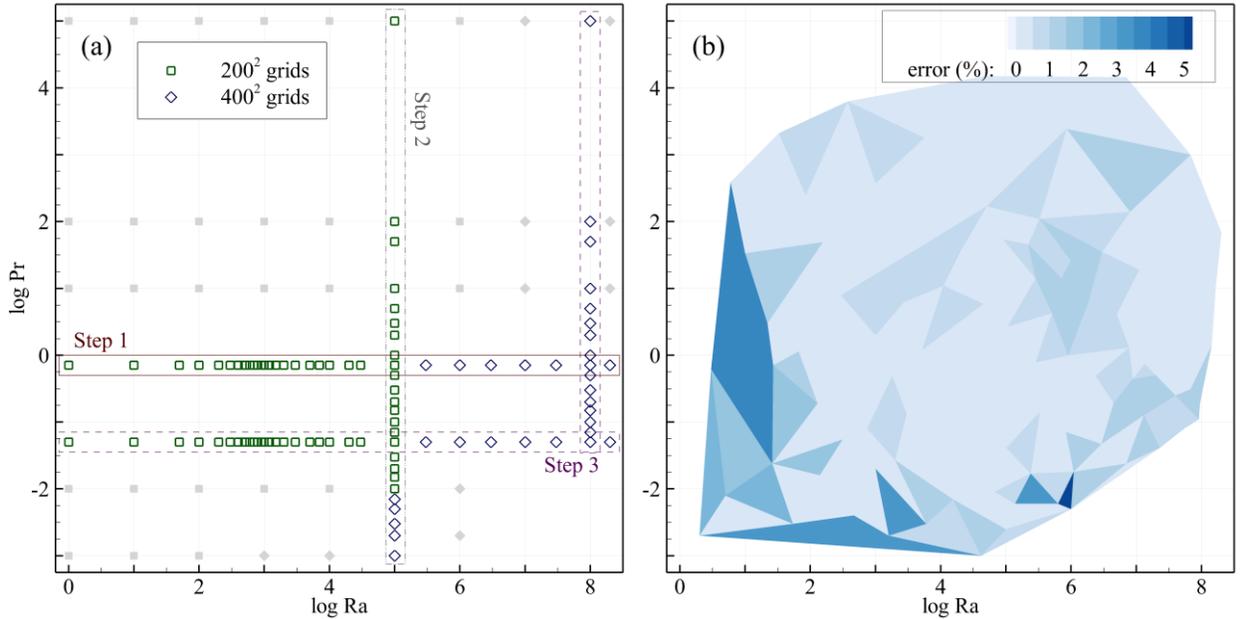

Fig. 5. a) The scatter graph for the training datasets for the different steps described in Table 3. b) the relative error contour for the test dataset applied to the final metamodel.

We also applied the above procedure to build a Nu metamodel for enclosures with a centered hole. We considered hollow enclosures with their inner walls being adiabatic while the other boundary conditions were the same as the original benchmark problem. A sample solution for this problem is presented in Appendix E. We generated a dataset of 798 new simulations for different inner widths ($d^* = 0.3, 0.5, 0.7, 0.8, 0.85, 0.9, 0.92, 0.95, 0.96, 0.97, 0.98$, and $0.99$) and combined it with the training data outlined in the previous section for $d^* = 0$. The case of $d^* = 0$ describes the original problem (with no hole at the center) and the case of $d^* = 1$ refers to a trivial case with no heat transfer domain (i.e., Nu = 0).

To generate our training dataset, we used a multi-grid approach based on the values of Ra and $d^*$. The dimensionless grid sizes for $d^*$ ranging from 0.3 to 0.9 were $1/160$, $1/320$, and $1/640$ for Ra $\leq 10^4$, $10^4 <$ Ra $< 10^7$, and Ra $\geq 10^7$, respectively. However, for larger inner widths, our model required finer grid sizes as small as $1/1600$. The scatter graph for the training dataset is presented in Fig. 6.

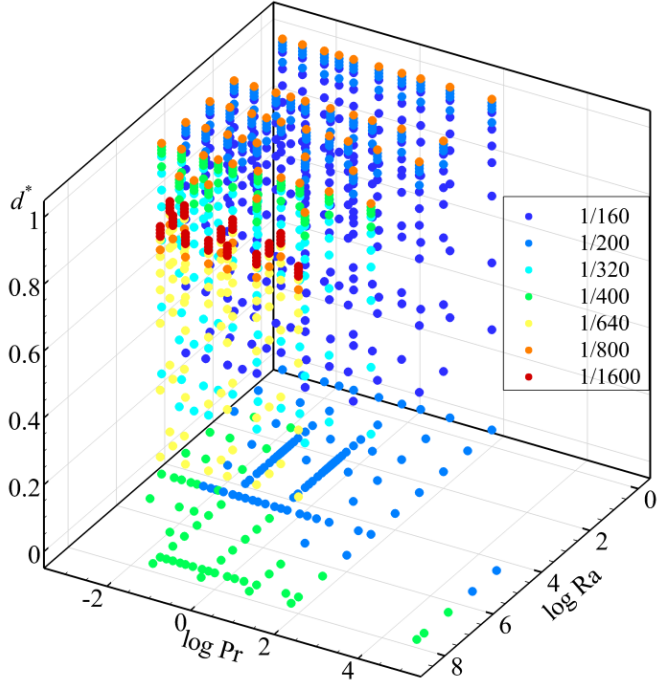

Fig. 6. The scatter graph for the training dataset for the case of hollow enclosures.

We employed the already trained DNN from the benchmark problem (Block III in Fig. 4) and transferred its learning to this new problem. The structure of the new DNN along with its training details are shown in Fig. 7. This structure is created by adding a new branch to account for variations of $d^*$ before combining all branches. Therefore, we froze the training on the two former branches (given that they were already trained), and only trained the new branch and the last two hidden layers (located after the concatenate layer, Fig. 7).

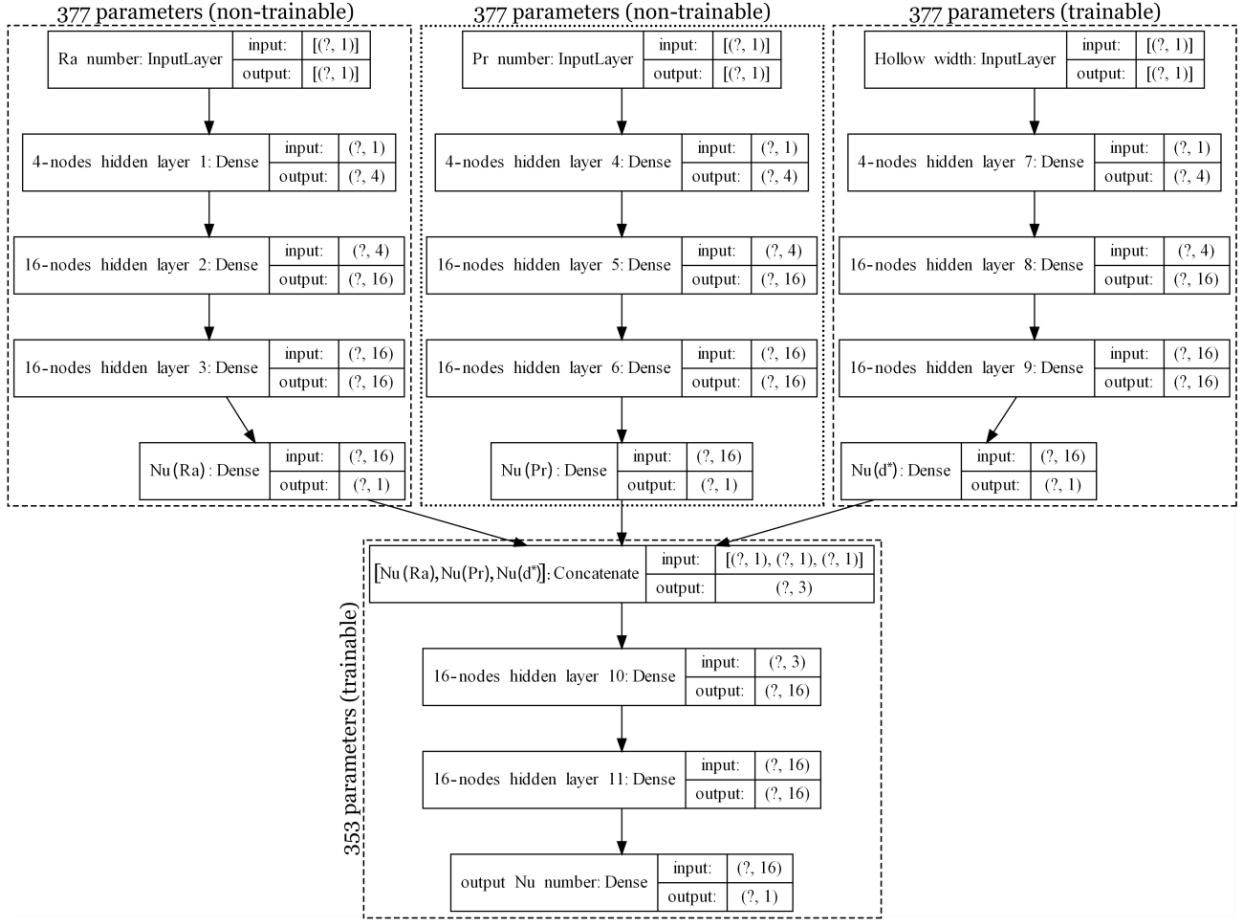

Fig. 7. The structure of the DNN for metamodeling Nu in hollow enclosures. Out of 1484 total parameters, 730 were trainable whereas the rest received their weights from the metamodel for the benchmark problem.

The training error for more than 20% of our data points was above 1%, which was higher than the maximum error resulting from an insufficient grid size. A further improvement in metamodel accuracy can be achieved by adding more data points wherever the training error is relatively high. The test result for our Nu metamodel for hollow enclosures is summarized in Table 4. For comparison, we also included the test result for an optimized model that did not benefit from TL, demonstrating that our TL approach resulted in an improved mean squared error of 63%. A scatter graph of predicted Nu values using the DNN metamodel versus our simulation data is presented in Fig. 8. The average and maximum differences between the predicted and simulated Nu results for the TL metamodel are 0.075 and 0.803, respectively. This test was carried out using a dataset of 200 simulations using the highest fidelity grid systems for $d^* = 0$, 0.2, 0.4, 0.6, 0.78, 0.82, 0.88, 0.91, 0.94, 0.965, and 0.985.

Table 4. The details of the test result for Nu metamodel for hollow enclosures.

|  | MSE ($/10^{-5}$) | MAE ($/10^{-3}$) | MRE (%) | SD (%) |
| --- | --- | --- | --- | --- |
| TL approach | 7.18 | 5.82 | 1.34 | 1.44 |
| Without TL | 19.46 | 8.08 | 1.86 | 2.59 |

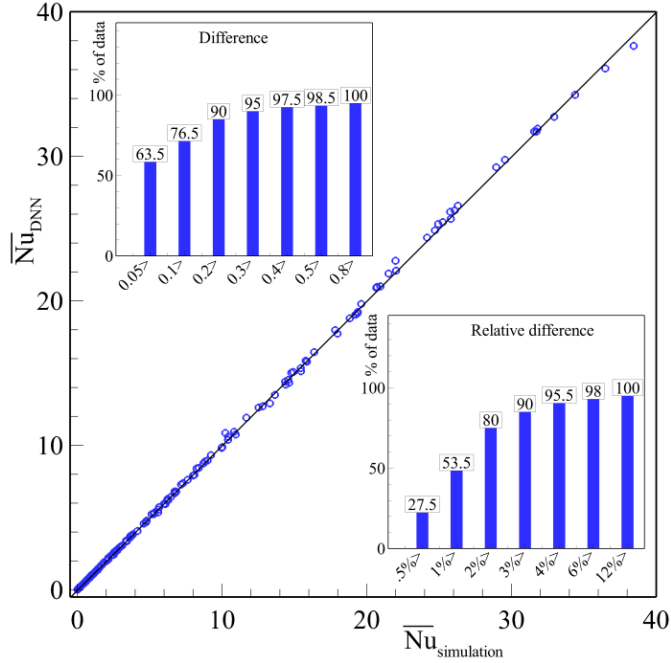

Fig. 8. The predicted Nu using the DNN metamodel versus simulation data for hollow enclosures.

## 4. Conclusion

While buoyancy drives natural convection, the performance of a natural convection system can be affected by other physical phenomena and different factors such as geometry, boundary conditions, and the behavior of a fluid. Data-driven metamodels that govern real-world natural convection systems need datasets that cover the entire feature space. Moreover, some unforeseen features may become active under a different process or after a system is redesigned. Among other limiting factors, generating a new full set of simulations or experiments is time-consuming. Our methodology using TL with DNNs can flexibly adapt to the expansion of the feature space when a natural convection system becomes more complicated and needs to be described more precisely. We considered a benchmark problem in square enclosures described by the Rayleigh and Prandtl numbers. We carried out a mesh refinement study on the input space, to find the appropriate grid size to perform accurate simulations. Although simulations using coarse grid systems may provide precise results, they only do so for a limited range of input parameters. For example, the higher the Ra, the finer the grid size required. As such, we utilized a multi-grid dataset for training our ANN in order to reduce simulation times. Any irregularity in the training loss could be an indication of inconsistency in the dataset due to grid error. We effectively denoised the dataset, and retrained the ANN based on abnormalities observed in training losses.

Secondly, we adopted a TL approach using DNNs to this problem. We demonstrated the capability of this approach in incorporating additional input features. We built a metamodel to predict Nu as a function of Ra (the only input variable), for an air-filled enclosure. We successfully applied a DNN and transferred its learning from an air-filled enclosure to an enclosure with arbitrary fluid (i.e., Pr was added as a new input). Furthermore, we extended the DNN to predict Nu in hollowed enclosures with different sizes. We conclude that our metamodel can be retrained to predict Nu for different natural convection problems. This TL strategy is versatile and can handle straightforward metamodeling problems for many different engineering systems.

**Appendix A: Mathematical Analysis for the Physical Model**

We modeled the natural convection flow in a square enclosure of width $D$ filled with a Newtonian fluid. The top and the bottom walls of the enclosure are adiabatic, and the left and right vertical walls are kept at a constant hot ($T_h$) and cold ($T_c$) temperature, respectively. The flow is laminar, incompressible and two-dimensional. Constant thermal properties are assumed except for density which is modeled by Boussinesq approximation. The governing conservation equations in the non-dimensional form can be expressed as Eqs. (A-1)-(A-4). In these equations, $u$ and $v$ are the velocity components in $x$ and $y$ directions, $P$ is the pressure, and $T$ is the temperature of the fluid. The superscript * is used to point out that the variables are in their dimensionless forms.

$$\frac{\partial u^*}{\partial x^*} + \frac{\partial v^*}{\partial y^*} = 0 \tag{A-1}$$

$$u^* \frac{\partial u^*}{\partial x^*} + v^* \frac{\partial u^*}{\partial y^*} = -\frac{\partial P^*}{\partial x^*} + \Pr\left(\frac{\partial^2 u^*}{\partial x^{*2}} + \frac{\partial^2 u^*}{\partial y^{*2}}\right) \tag{A-2}$$

$$u^* \frac{\partial v^*}{\partial x^*} + v^* \frac{\partial v^*}{\partial y^*} = -\frac{\partial P^*}{\partial y^*} + \Pr\left(\frac{\partial^2 v^*}{\partial x^{*2}} + \frac{\partial^2 v^*}{\partial y^{*2}}\right) + \Pr \operatorname{Ra} T^* \tag{A-3}$$

$$u^* \frac{\partial T^*}{\partial x^*} + v^* \frac{\partial T^*}{\partial y^*} = \left(\frac{\partial^2 T^*}{\partial x^{*2}} + \frac{\partial^2 T^*}{\partial y^{*2}}\right) \tag{A-4}$$

The non-dimensional variables in the above equations are defined as $x^* = x/D$, $y^* = y/D$, $u^* = u/(\alpha/D)$, $v^* = v/(\alpha/D)$, $P^* = P/(\rho \alpha^2/D^2)$, $T^* = (T - T_c)/(T_h - T_c)$, where $\alpha = k/(\rho c)$ is the thermal diffusivity of fluid. We define Pr and Ra as

$$\Pr = \frac{\nu}{\alpha}, \qquad \operatorname{Ra} = \frac{g\beta(T_h - T_c)D^3}{\nu\alpha} \tag{A-5}$$

where $g$ is the gravitational acceleration; $\nu, \rho, c, k$, and $\beta$ are kinematic viscosity, density, specific heat, thermal conductivity, and the volumetric coefficient of thermal expansion, respectively.

The hydrodynamic and thermal boundary conditions are specified in Eqs. (A-6) and (A-7), respectively.

$$u^* = 0 \text{ and } v^* = 0 \text{ (for all walls)} \tag{A-6}$$

$$T^* = 1 \text{ (on left wall)}, \qquad T^* = 0 \text{ (on right wall)}, \tag{A-7}$$

$\partial T^*/\partial y^* = 0$ (on top and bottom walls),

The local and average Nu on the left ($X = 0$) and right ($X = 1$) walls are defined as

$$\text{Nu}_y = \left|\frac{\partial T^*}{\partial x^*}\right|_{x^*=X}, \qquad \overline{\text{Nu}} = \int_0^1 \left|\frac{\partial T^*}{\partial x^*}\right|_{x^*=X} dy^* \tag{A-8}$$

where $\overline{\text{Nu}}$ has the same value at both the vertical walls, due to the steady-state condition.

**Appendix B: Numerical Method and Validation**

The coupled governing Eqs. (A-1)−(A-4) were transformed into algebraic equations using the finite volume method, and the SIMPLE algorithm was used for pressure-velocity coupling for the momentum equations. Our numerical code was validated for the case of an air-filled enclosure, and showed agreement with published results. More details on the numerical method and its verification can be found in our previous papers [29-31]. The average Nu and the maximum dimensionless stream function ($\psi^* = \psi/\alpha$) were evaluated for different grid systems at different Pr (for the highest Ra), as summarized in Table B-1. The tested grids are all uniform and structured whereas a boundary mesh is applied in which the cells adjacent to the walls were split in half to account for higher gradients near the walls. In comparison to the results obtained by an 800×800 grid system, the highest errors for the results obtained using the 200×200 and 400×400 grid systems were about 1 and 0.5 percent, respectively (for the most stringent cases).

Table B-1. Comparison of the average Nu and the maximum stream function for $\text{Ra} = 10^8$ for different grid sizes; the values in parentheses are the relative errors compared with the most accurate results.

| Grid system | Pr = 0.1 | | Pr = 1 | | Pr = 10 | |
|---|---|---|---|---|---|---|
| | $\overline{\text{Nu}}$ | $\psi^*_{max}$ | $\overline{\text{Nu}}$ | $\psi^*_{max}$ | $\overline{\text{Nu}}$ | $\psi^*_{max}$ |
| 25×25 | 28.94 (14.8%) | 56.06 (31.8%) | 34.65 (12.4%) | 63.46 (11.7%) | 35.64 (12.1%) | 68.29 (9.44%) |
| 50×50 | 26.78 (6.27%) | 46.86 (10.2%) | 33.63 (9.12%) | 59.15 (4.12%) | 35.07 (10.3%) | 66.19 (6.07%) |
| 100×100 | 24.79 (1.65%) | 44.94 (5.65%) | 31.05 (0.74%) | 57.77 (1.69%) | 32.19 (1.27%) | 63.96 (2.50%) |
| 200×200 | 25.09 (0.43%) | 42.84 (0.73%) | 31.07 (0.82%) | 57.04 (0.40%) | 32.13 (1.06%) | 62.72 (0.51%) |
| 400×400 | 25.14 (0.26%) | 42.47 (0.15%) | 30.93 (0.34%) | 56.90 (0.15%) | 31.95 (0.51%) | 62.51 (0.18%) |
| 800×800 | 25.20 | 42.53 | 30.82 | 56.81 | 31.79 | 62.40 |

**Appendix C: The multi-grid simulation**

We show that the grid dependency of the solution depends on the value of Ra and Pr. Figure C-1 demonstrates Nu as a function of Ra and Pr. As can be seen in the constant-Pr curve of Fig. C-1a, a 25×25 grid system reliably simulates the problem for Ra up to $10^3$ (having a maximum difference of 0.2% relative to the results obtained from a 200×200 grid system). Likewise, a 50×50 grid system predicts Nu accurately enough provided that Ra $< 10^5$ (with a maximum relative difference of below 0.6% in comparison to the result obtained using a 200×200 grid system). The validity of this statement was also assessed for different Pr, as presented in Fig. C-1b for Ra $= 10^5$. We conclude that we can rely on the above statement except for cases with low Pr for which coarse grid systems result in high error values (due to convergence difficulties). Therefore, we employ models having 25×25 and 50×50 simulation grids for Ra $\leq 10^3$ and Ra $< 10^5$, respectively, and a 200×200 grid system for other cases. Nonetheless, to maintain the accuracy of the Nu result above 99%, we employ a 400×400 grid system for low Pr or high Ra (as illustrated in Fig. 1a). Our analysis provides an approximate criterion for selecting grid sizes for different input ranges. Nonetheless, based on the errors between the original data and the ANN model, we further revised the data points by using finer grid sizes.

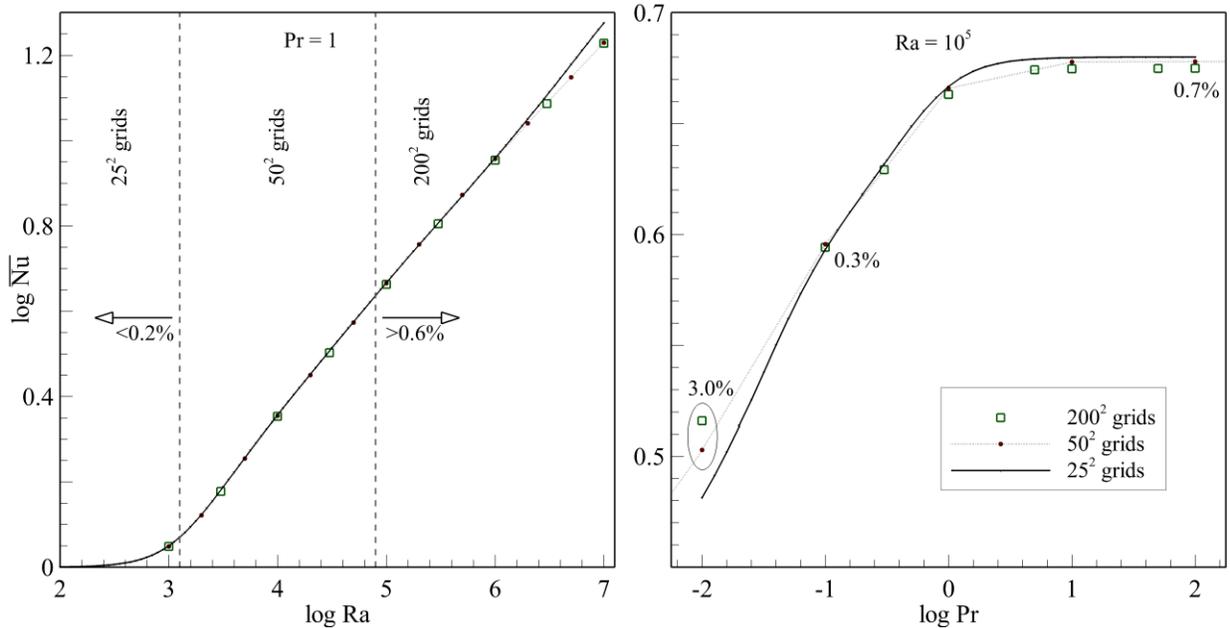

Fig. C-1. Nu as a function of a) Ra at a fixed Pr $= 1$, and b) Pr at a fixed Ra $= 10^5$.

**Appendix D: ANN Optimization**

We optimized the ANN architecture using the original training dataset. The dataset (480 simulations) was first split into training and validation sets with a ratio of 15% (408 training data and 72 validation data) using the model selection function in the Scikit-learn library (the test dataset was not used in the ANN optimization process). We trained our ANNs using the Keras 2.3.1 library with TensorFlow 2.2.0 as the backend. We used the Hyperband algorithm [33] that is a Bandit-based approach to optimize the hyperparameters of our model. Our neural networks consisted of a hidden layer with 4 nodes next to the input layer, and a combination of 1 to 5 intermediate hidden layers with either 4, 8, 16, or 32 nodes within every hidden layer. As Fig. C-1 suggests, 'tanh' is a good approximator for the prediction of Nu; thus, we applied the 'tanh' activation function for all the layers of our ANNs. We tested the Adam optimizer [34] with different learning rates from 5×10-4 to 10-3 with "log" sampling (which assigns equal probabilities within the range of each order of magnitude) in the Keras HyperParameters container. After testing different combinations through the Hyperband algorithm, a structure as described in Table D-1 was selected.

Next, we ran a limited number of brute-force tunings on the optimized model using more epochs. At this stage, we applied a multi-stage training procedure using different batch sizes of $2^{n+1}$ applied sequentially ($n$ = 1, 2, …, 7) and picked the best model among the models produced after each of these stages. Sequentially changing the batch size enabled us to find a well-trained ANN without worrying about having a proper value for the batch size as well as the learning rate, which changes adaptively for the Adam optimizer. Nonetheless, we tested a limited number of batch sizes from 0.0001 to 0.001 and performed the multi-stage training separately. No dropout layer was required, and a learning rate of 0.0005 provided the lowest validation loss. One may validate this approach on another benchmark dataset (for example, the one generated by Lejeune [35]) to compare its efficiency against other training procedures.

Table D-1. The structure of the optimized ANN for the Nu prediction.

| | |
|---|---|
| ANN class: | multilayer perceptron (MLP) feedforward |
| Layers: | input: 4 nodes, 2 inputs<br>2 hidden layers: 16 nodes each<br>output layer: 1 node, 1 output |
| Total parameters: | 381 |
| Activation function: | 'tanh' $(= 2/(1 + e^{-2x}) - 1)$ |
| Optimizer: | Adam |
| Learning rate: | 0.0005 |
| Batch size: | $2^{n+1}; n = 1,2,...,7.$ |

**Appendix E: The hollow enclosure**

We simulated hollow enclosures with adiabatic inner walls. The details of the numerical method and its verification can be found in our previous paper [31]. To account for the grid size dependency, we compared the results of the average Nu as well as the maximum stream function for different grid sizes. The dimensionless temperature and stream function results for a simulation conducted for an air-filled hollow enclosure are presented in Fig. E-1. In comparison with the original enclosure, hollow enclosures do not necessarily have lower heat transfer performance. Even though the heat transfer domain is smaller, a narrow passage brings higher velocity gradients, and therefore, this combined effect may produce higher overall Nu. Thus, at fixed Ra and Pr, there exists a non-zero inner width ($d^*$) for maximum Nu. For example, for air-filled enclosures, a $d^* = 0$ enclosure has the highest Nu at Ra = $10^3$. However, as Ra increases, the maximum Nu is seen at larger $d^*$ (e.g., $d^* \cong 0.5$ for Ra = $10^5$ and $d^* \cong 0.85$ for Ra = $10^8$). Our Nu metamodel can be exploited to provide these optimum inner widths for all Ra and Pr.

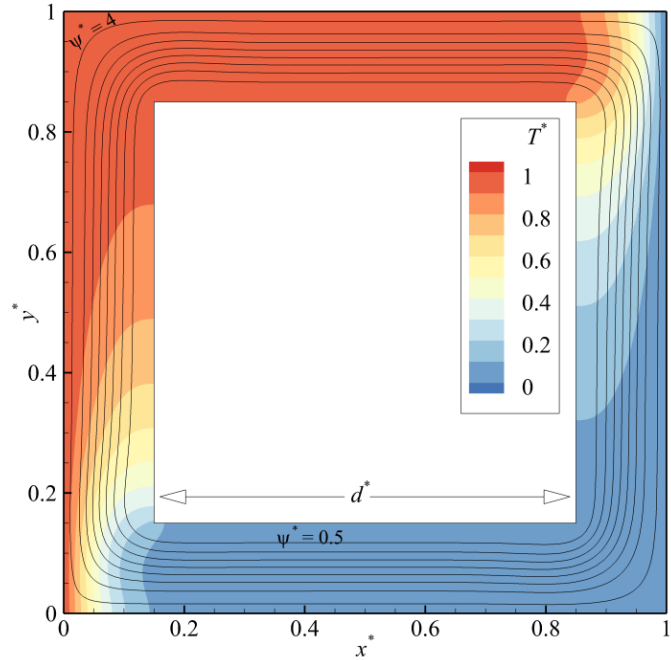

Fig. E-1. The simulation result for an air-filled hollow enclosure ($d^* = 0.7$) at Ra = $10^5$; The lines show contours of streamlines, while the temperature contours are represented by colors.

**Supplementary material**

An open-source version of this paper is available at https://github.com/engdatasci/NCTL. The codes, datasets, and resulting metamodels can be found in this repository.